\documentclass[sigchi]{acmart}

\AtBeginDocument{%
  \providecommand\BibTeX{{%
    \normalfont B\kern-0.5em{\scshape i\kern-0.25em b}\kern-0.8em\TeX}}}

\setcopyright{none} 
\usepackage{graphicx}
\usepackage{amsmath}
\begin{document}

\renewcommand\footnotetextcopyrightpermission[1]{}
\pagestyle{plain} 
\settopmatter{printacmref=false}
\title{Lessons Learned Developing and Extending a Visual Analytics Solution for Investigative Analysis of Scamming Activities}

\author{Ronak Tanna}
\email{rtanna@asu.edu}
\affiliation{%
  \institution{Arizona State University}
}

\author{Shivam Dhar}
\email{sdhar3@asu.edu}
\affiliation{%
  \institution{Arizona State University}
}

\author{Ashwin Sudhir}
\email{asudhir1@asu.edu}
\affiliation{%
  \institution{Arizona State University}
}

\author{Shreyash Devan}
\email{srdevan@asu.edu}
\affiliation{%
 \institution{Arizona State University}
 }

\author{Shubham Verma}
\email{sverma41@asu.edu}
\affiliation{%
 \institution{Arizona State University}
 }

\maketitle

\section{Introduction}
Cybersecurity analysts work on large communication datasets to perform investigative analysis by painstakingly going over thousands of email conversations to find potential scamming activities and the network of cyber scammers. Traditionally, experts used email clients, database systems and text editors to perform this investigation. With the advent of technology, elaborate tools that summarize data more efficiently by using cutting edge data visualization techniques have come out. \textit{Beagle} [1] is one such tool which visualizes the large communication data using different panels such that the inspector has better chances of finding the scam network. 

This paper is a report on our work to implement and improve the work done by Jay Koven et al. [1]. The original paper describes a tool called Beagle, which was developed by the authors to analyze how e-mail scammers interact with each other and their victims.

For the creation of this tool, they worked with a security company called Agari, who was looking for a tool that would help them analyse the e-mail scamming activities. While working with Agari they found that even though experienced analysts were very proficient at going through these e-mails using numerous tools to aid them, they could greatly benefit from a visualization tool. This benefit is in the form of query reformulation and content summarizing through visual representation, which should help the analysts not only be more efficient but allow for new types of analyses.
On top of the ideas presented in the original paper, we have implemented a few more visualizations that we feel would help in grouping and analyzing the e-mail data more efficiently. Lastly, we have also presented a case study that shows the potential use of our tool in a real-world scenario.

\section{Visualization Design}

The visual analytics solution - Beagle, was built for investigative analysis of scamming activities on communication datasets consisting of unstructured text, social network information, and metadata. The system provides capabilities such as a progressive and reversible data query interface with similarity to email clients, coordinated views to keep results in context defining who, what and when, provision to make queries visible and easy to extend or modify, content reduction and extraction methods, content summarizing techniques for email data, and capability of tagging data points to externalize knowledge via different panels as discussed in the sections below and shown in figure 1.

\subsection{Interactive query panel} It provides users with a filtering capability based on metadata in the dataset such as subject and content. Option to add or remove the query terms on the fly adds to the usefulness of the feature. Based on the filtered data, all other panels get updated.

There are control buttons that help in uploading a dataset to be analyzed and downloading the sequence of actions performed by the user. These are marked as 2 in figure 1.

\subsection{Correspondent panel} This panel displays email ids of the people involved in email exchanges in the filtered dataset. It also shows the number of emails received and sent by these correspondents in form of a pie chart. These statistics help analysts to find out the spammers based on their activity, for example a correspondent who might be sending a lot of emails could be a potential spammer.  

\subsection{Contact graph panel} Built as a modal controlled through a button click - refer btn1 in figure 1, it shows a graphical representation of correspondents involved in the email exchanges for the filtered dataset, which can be seen in the panel tagged 8 in the figure. A button to remove edges and nodes is provided to reduce the density of the network and drill down to a specific sub-network. Similarly, a button to add edges and nodes based on the sequence of deletions helps to create the dense network.

\begin{figure*}
\includegraphics[width=0.9\textwidth]{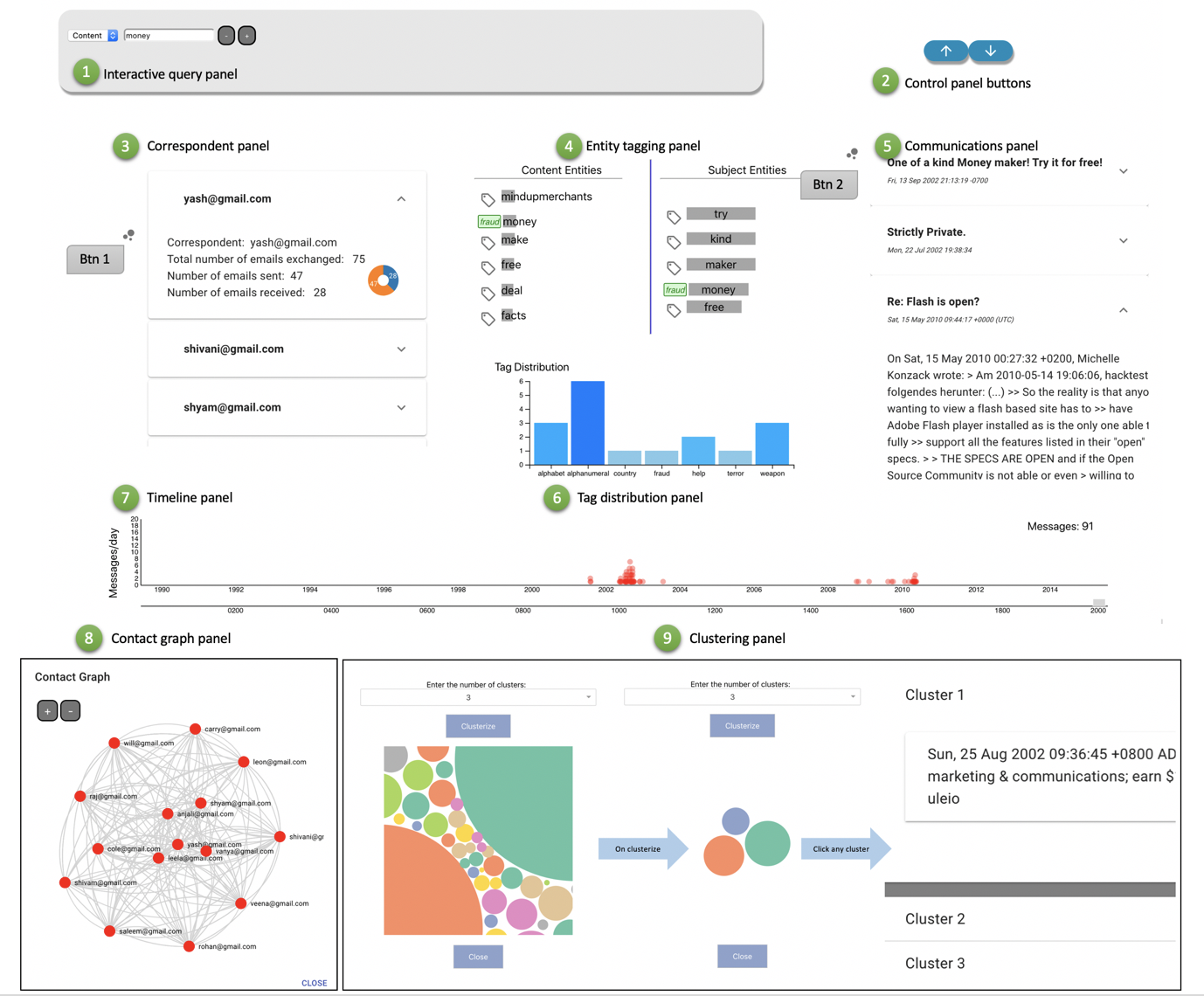}
\caption{The different data exploration panels in Beagle.}
\end{figure*}

\subsection{Entity tagging panel} Term Frequency - Inverse Document Frequency (TF-IDF), an information retrieval technique is used to find interesting entities from the email subject and content and is shown in the entities panel, tagged as 4 in figure 1.

\begin{equation}
tf-idf(t,d,D) = f_{(t,d)} * log (|D|/|{d \in D : t \in d}|)
\end{equation}
Here t is a term in the document, d and D stands for the collection of documents.

The words are sorted in the decreasing order of TF-IDF values represented by the width of the bar. A user can also add new/existing tags to the entities of interest using the context menu option. These tags help users to identify the pattern in the filtered results and also helps in understanding the commonality of spamming activities across the different datasets. 

\subsection{Communications panel} The emails that are part of the filtered data are shown here in a way similar to the email clients, with the option of expanding the subject header to look through the content, as shown by the tag 5 in figure 1. The header shows subject along with datetime if present in the dataset whereas the content displays the email content in plain text including receiver and sender information.

\subsection{Timeline panel} All the emails exchanged between the time period as specified by the filtered data are shown here in the form of an interactive plot tagged as 7 in figure 1. A slider option is provided to zoom in to a specific day, month or year to find the exact number of email exchanges.

\section{Extension}

We propose extensions to the system in order to gain more insights about the data by providing additional capabilities that could help an analyst to investigate further. We also make some changes in the user interface to improve the usability and overall user experience.

\subsection{Clustering panel}
It provides users with an option to form clusters based on content similarity via btn2 as shown in figure 1. Initially, all the emails are shown in a packed layout form, and once a user clicks on \textit{clusterize} button after selecting the number of clusters from the drop-down option, the cluster heads are shown. A click on these cluster heads shows the emails that are part of each cluster. The sequence of actions is shown in the section tagged as 9 in figure 1.

\subsection{Tag Distribution panel}
The panel displays the number of tags being assigned to the entities of interest in the form of a color varying histogram. It allows a tooltip-based interaction and gets dynamically updated as a user assigns a tag to an entity from the entity panel. It allows tracking of the tagged knowledge points across different datasets.

\subsection{UI Component styling}
For improving the overall user experience of the system, we chose to style the components to make them look aesthetically pleasing.

The correspondent panel was modified to use expansion panel-based design that allows a user to expand the correspondent header to look for more details such as the pie chart and email exchanges. 

The entity tagging panel was improvised to display the tags assigned in a more distinctive and compact way. Tagging functionality doesn't seem to work as expected in the original system.

The contact graph panel now renders a fluid graph with a set of interactions like highlighting an edge and the corresponding nodes connected by the edge, capability of reducing the graph density by either removing edges or nodes, etc.

The expansion panel was reused for rendering the communications panel to make it more compact and sleek.

Timeline panel has been modified from a histogram to a scatter plot with a zooming slider functionality for year, month and day.

\section{Case Studies}
For privacy reasons, emails of the users are not provided in the dataset. We synthetically add emails in the datasets from a pool of selected emails in order to run visualizations on the dataset. This is also required to present case studies showing the effectiveness of the tool built.

\subsection{Dataset I}
This section describes a case study on a spam emails dataset, dataset I [2]. We search for words often seen in spam emails such as \textit{click} and \textit{link}. On searching for \textit{click} as a content query term, we could see the email exchanges in the time panel as a uniform distribution over a period of time - 2000 to 2010 with the maximum of 18 emails/day. On adding \textit{link} to the query, the maximum email exchanges came down to 6/day with a distribution shared between 2002 to 2003 and 2009 to 2011. Further, by adding subject query as \textit{spam} we could see a huge concentration of email exchanges only in 2009 with shivani@gmail.com as the correspondent with the largest number of email exchanges.

\subsection{Dataset II}
This section describes a case study on a fraudulent email corpus, dataset II [3]. The user interacts with the system through the query interface, where he enters the content as \textit{money}. The timeline panel shows the number of email exchanges as a distribution between 2003-2008. Tags are assigned to the words such as \textit{receipt, goods, company, inventory, officials} are assigned a politics tag whereas entities like \textit{urgent, business, dollars, money} are assigned a suspicious tag. Another filter with content as \textit{transfer} is added, which reduces the dataset analyzed. The user can see the tags for the words assigned earlier if found in the entities panel as subject or content. As filters are added, the dataset under consideration narrows down further giving a clear idea of a pattern if any in the spam emails. Filters like content as \textit{Nigeria}, content as \textit{urgent}, subject as \textit{urgent}, content as \textit{bank} are added, which makes 2003 and 2007 as the years with maximum number of email exchanges. The correspondent panel always shows the person with the maximum number of email exchanges at the top, this helps in identifying an anomaly or a spammer out of all the correspondents. The clustering panel helps in segregating the emails into different buckets, for instance above query gave two sets of clusters - one with \textit{urgent}in the subject and another one with \textit{very urgent} in the subject. 

The addition of synthetic emails did help in visualizing the attributes to a great extent but the case studies couldn't get a closure due to the lack of correlation between the correspondents and the entities referred in the email content.  
\section{Future Work}

We envision the project to be extensible in various ways. The potential directions in which promising future work can take place include:
\begin{itemize}
    \item Multi user option with the ability for each user to upload and save multiple data sets. This would enable an investigative analyst to work on multiple cases at the same time.
    \item The ability to search the tagged words and to automatically tag emails that contain those words.
    \item Manually choosing cluster centers around which the clustering happens. This could mean that given an email that asks about credit card details as part of a scam that the user knows about and a normal communication email, the user, after choosing those as cluster centers should be able to get all similar emails in the two respective clusters to be able to visualize neatly.
    \item Integration with a live mailbox that gives an investigative analysis user full fledged power to do everything that he/she does with the mailbox on a daily basis in a smooth way via the application itself.
\end{itemize}

We understand that each component's necessity can only be determined by a user survey that confirms it; we only list the most promising extensions to our system and leave it to future research to drill down on specifics.




\section{References}

\bibliographystyle{ACM-Reference-Format}
 [1] Koven J, Felix C, Siadati H, Jakobsson M, Bertini E. 2018. Lessons Learned Developing a Visual Analytics Solution for Investigative Analysis of Scamming Activities. IEEE Transactions on Visualization and Computer Graphics ( Volume: 25 , Issue: 1 , Jan. 2019 ). DOI: 10.1109/TVCG.2018.2865023\newline
[2] www.kaggle.com/c/adcg-ss14-challenge-02-spam-mails-detection\newline
[3] Radev, D. (2008), CLAIR collection of fraud email, ACL Data and Code Repository, ADCR2008T001, http://aclweb.org/\newline

\end{document}